\documentclass[12pt,showpacs,preprintnumbers,amsmath,amssymb]{revtex4}
\usepackage{epsf}
\usepackage{here}
\usepackage[dvips]{graphicx}
\usepackage{amsmath}
\usepackage{amssymb}

\def\ds{\displaystyle}

\newcommand{\pdl}[2]{\frac{\partial #1}{\partial #2}}
\def\infint{\int_{-\infty}^\infty}
\def\tdf{\widetilde{\delta f}}
\def\tdk{\widetilde{\delta f_k}}
\def\vek{\varepsilon_k}

\begin{document}
\baselineskip 24pt

\title{Formation of fractal structure in many-body systems
with attractive power-law potentials}
\author{
  Hiroko Koyama\footnote{koyama@gravity.phys.waseda.ac.jp}}
\affiliation{Department of Physics,
Waseda University, Shinjuku, Tokyo, 169-8555, Japan
}
\author{Tetsuro Konishi\footnote{tkonishi@r.phys.nagoya-u.ac.jp}}
\affiliation{Department of Physics, 
Nagoya University, Nagoya, 464-8602, Japan}

\date{\today}

\begin{abstract}
We study the formation of fractal structure in one-dimensional
many-body systems
with attractive power-law potentials.
Numerical analysis shows that 
the range of the index of the power for which
fractal structure emerges is limited.
Dependence of the growth rate  on wavenumber and power-index is obtained
by linear analysis of the collisionless Boltzmann equation, which 
supports the numerical results.
 \end{abstract}
\maketitle
\section{Introduction}
Formation of spatial structures is an interesting and important phenomenon
in nature.
It is seen over a wide range, from
protein folding in biological systems~\cite{Fersht-1998,Karplus-2000} to
large-scale structure in the universe~\cite{Peebles-book-1980}.
The theoretical origins of such structures are quite important
and will be classified into several classes.
One of the most interesting areas within the field of dynamical systems is that
some remarkable structure and organization is created dynamically 
by the mutual interaction among the elements~\cite{dauxois-2000,barre-2001}.

Recently we have discovered that spatial structure with fractal distribution
emerges spontaneously from
 uniformly random initial conditions
in a one-dimensional self-gravitating system, that is
the sheet model~\cite{hktk-1}.
What is noteworthy in this phenomenon is that the spatial structure is not
given at the initial condition, but dynamically  created from a state
without spatial  correlation.
Succeeding research clarified that the structure is created first in 
small spatial scale then grows up to large scale
through hierarchical clustering~\cite{hktk-2}, and the structure is 
transient~\cite{hktk-3}.
It is quite interesting that 
some remarkable spatial structures are emerged instead of monotonous thermal
relaxation in Hamiltonian system.
%




The emergence of fractal structure 
is a typical example that 
systems of many degrees of freedom
are self-organized by dynamics themselves.
Hence to clarify its dynamical mechanism
 is very important subject toward
understanding physics of self-organization
of matter. 

A way to clarify the dynamical mechanism is to know which class of pair 
potential can form the fractal structure. 
Here we note an important fact that 
fractal structure does not have characteristic spatial scale, nor does
the potential of the sheet model,
since the pair potential is power
of the distance.
Hence the scale free property of potential
may be a key-stone to understand
the dynamical mechanism.
The question is if the fractal structure can be formed in not only
the sheet model, but also other systems with power-law potentials.

In this paper we study 
the possibility that the fractal structure can be formed in 
more general systems without characteristic spatial scale
which is  extended from sheet model.
Here we adopt the model as the system with attractive power-law
 potentials.
At first we examine the formation of the fractal structure
by numerical simulation for various values of the power index of the potential.
Next we perform linear analysis to consider  the numerical results.

In Sec.\ref{sec:model} we introduce many-body systems with power-law
potentials.
In Sec.\ref{sec:sheet} we review the formation of power-law correlation 
in the sheet model.
In Sec.\ref{sec:simulation} we carry out numerical simulation.
In Sec.\ref{sec:analysis} we analyze linear perturbation of the
collision-less Boltzmann equation.
The final section is devoted to summary and discussions.

\section{Model}
\label{sec:model}

We consider the model where
many particles with an uniform mass interact with purely attractive pair potential
of power-law, which is described by the Hamiltonian \cite{MPT}
\begin{equation}
 H=K+U=\sum_{i=1}^{N}\frac{p_i^2}{2}
+\sum_{i=1}^{N-1}\sum_{j>i}^{N} 
\left|x_i-x_j\right|^{\alpha},
\label{eq:model}
\end{equation}
where $x_i$ and $p_i$ are the position and momentum a particle, respectively. 
The first term is kinetic energy and the second term is potential energy.
For simplicity in this paper we consider the system where motion of
particles is bounded to one-dimensional direction.

For the special case $\alpha=1$,
the Hamiltonian (\ref{eq:model})
applies to a system of $N$ infinite parallel mass sheets, where each
sheet extends over a plane parallel to the $yz$ plane and moves along
the $x$ axis under the mutual gravitational attraction of all the other
sheets. 
The Hamiltonian of the sheet model
~\cite{hohl-feix-1967,rybivki-1971-apss,severne-1984-apss,Miller-2} 
is usually written in the form
\begin{equation}
 H=\sum_{i=1}^{N}\frac{p_i^2}{2}
+\sum_{i=1}^{N-1}\sum_{j>i}^{N} \left|x_i-x_j\right|.
\label{eq:model-sheet}
\end{equation}

In the previous letters~\cite{hktk-1,hktk-2,hktk-3} we investigated the time evolution
of the sheet model (\ref{eq:model-sheet}) to show that fractal structure
emerges from non-fractal initial conditions. In the Secs. 
\ref{sec:simulation} and \ref{sec:analysis}
we will clarify the relation between the power index of the pair potential
$\alpha$ and 
the formation of the fractal structure by
investigating time evolution of more general
model (\ref{eq:model}) numerically and analytically.

\section{formation of fractal structure in the sheet model ($\alpha =1$)}
\label{sec:sheet}
Before we investigate the time evolution of the model (\ref{eq:model}), 
we briefly review our previous works ~\cite{hktk-1,hktk-2,hktk-3}
on the model (\ref{eq:model-sheet}).
We found that fractal structure is formed
in the sheet model~(\ref{eq:model-sheet})
from uniformly-random initial conditions~\cite{hktk-1,hktk-2}.
In Figure~\ref{fig:corr-and-xu-rnd-003} we show a typical 
example of such fractal structure  and a snapshot of the $\mu$-space 
configuration.

\section{Numerical simulation}
\label{sec:simulation}

In this section we examine by computer simulations 
if the fractal structure is formed 
in the systems (\ref{eq:model}).
Here we carry out the case $\alpha\ge 1$ to avoid
divergence when two particles come close.

The non-fractal initial conditions where
fractal structure is
emerged are characterized typically by those of virial ratio 
$2E_{kin}/ E_{pot} = 0$ in the sheet model~\cite{hktk-1}.
(Spatial distribution is set to be random.) These state of zero
velocity dispersion corresponds to the limiting case of zero 
thermal fluctuation, and is called cold-random condition.
Therefore we adopt this cold-random initial condition to investigate
the time evolution for various value of $\alpha$.
In our simulation we use 4-th order of the symplectic integrator with a fixed time step
$\Delta t=2\pi/10^{4}$ and $N=65536$ particles.
In what follows we show numerical results for 
two typical examples: $\alpha=1.125$ and 
$\alpha=1.5$. For other values of $\alpha$  behavior of the systems
varies gradually in accordance with the change of $\alpha$.

\subsection{The case $\alpha=1.125$}
At first, we consider the case
the interaction force deviates slightly from sheet model.
Here we show the numerical results in the case $\alpha=1.125$
in Figs.\ref{fig:11mu} and \ref{fig:11xi}.
We display particle distribution in $(x,v)$ space ($\mu$-space) 
in Fig.\ref{fig:11mu}.  
In the course of time evolution 
we see that many whirlpools nest in the $\mu$-space to form 
the hierarchical structure.
In Fig.\ref{fig:11xi} we show a box counting dimension of the spatial distribution.
We can see that 
the dimension is $D\simeq 0.83$ (Fig.\ref{fig:11xi}). 
Therefore our numerical results suggest that the fractal structure is formed.
These behaviors are similar qualitatively
to the sheet model, $\alpha=1$ \cite{hktk-1}.
We find that fractal structure can be formed as well as the sheet model.

\subsection{The case $\alpha=1.5$}
Next, we consider the interaction deviate further from sheet model.
Time evolution change qualitatively as $\alpha$ increases.
Here we show the numerical results in the case $\alpha=1.5$.
We display particle distribution in $\mu$-space
in Fig.\ref{fig:15mu}, and in Fig.\ref{fig:15xi} we show a box counting
dimension for spatial distribution.
Differently from the case $\alpha=1.125$, a single spiral is rolled up in $\mu$-space.
Therefore  fractal structure is not formed.

We can summarize these numerical results in this section
that the fractal structure can be constructed as well as the  sheet model,
but range of $\alpha$ for which the fractal structure is created
is limited;
it can not be constructed for large value of $\alpha$.

\section{Analysis of linear perturbation}
\label{sec:analysis}
In this section we clarify the physical reason analytically
why fractal structure can not be constructed for the large value of the
power index of the
potential in numerical simulation in the previous section.
The formation of the fractal structure occurs at the relative early stage in
the whole-evolution history\cite{hktk-3}.
Then it is instructive for understanding the mechanism by which the
structure is formed to know the qualitative properties 
of the short-term behaviors by linear analysis.

In this section we derive the dispersion relation from the collisionless Boltzmann
equation (CBE), which describes the growing rate of the linear
perturbation~\cite{Galactic-Dynamics,Inagaki-Konishi-1993}.

\subsection{Collisionless Boltzmann equation}
CBE is defined by
\begin{equation}
\left\{
  \pdl{}{t}  + p\pdl{}{x}
+\infint dx' F(x-x')\left( \infint dp' f(x',p',t)\right)\pdl{}{p}
\right\} f(x,p,t)= 0,
\label{eq:cbe}
\end{equation}
where $F$ is 2-particle force which is related
with a pair-potential $U$ by
\begin{equation}
F(x) = -\frac{dU}{dx},
\end{equation}
and $f$ is the one-particle distribution function.
For simplicity we consider the system is extended in infinite region
$-\infty<x<\infty$.
It is clear that the state of uniform spatial density with an arbitrary
velocity distribution,
\begin{equation}
f(x,p,t) = f_0(p),
\label{eq:statio}
\end{equation}
is a stationary state. We impose the following small perturbation over
the stationary state (\ref{eq:statio})
\begin{equation}
f(x,p,t) = f_0(p) + \delta f(x,p,t).
\end{equation}
The linearized equation for $\delta f$ is
\begin{equation}
  \left( \pdl{}{t} + p\pdl{}{x}\right) \delta f(x,p,t)
= -\infint dx' F(x-x') \infint dp' \delta f(x',p',t) \pdl{}{p}f_0(p).
\label{eq:cbe-l}
\end{equation}

Now  we define the Fourier-Laplace transform by that
$x$ is Fourier transformed and $t$ is Laplace transformed, that is
\begin{equation}
  \tdf(k,p,\omega) \equiv
\int_0^\infty dt e^{-i\omega t}\infint dx e^{ikx}\delta f(x,p,t).
\end{equation}
Fourier transform is
\begin{equation}
\widehat{\delta f}(k,p,t)\equiv \infint dx e^{ikx}\delta f(x,p,t),
\end{equation}
and
\begin{equation}
\hat F(k) \equiv \infint dx e^{ikx} F(x).
\end{equation}

Then the Fourier-Laplace transformed equation of Eq.(\ref{eq:cbe-l}) is
\begin{equation}
\vek(\omega)\tdk(\omega) 
= \frac{-1}{{i( -\omega + kp)}}\widehat{\delta f}(k,p,0),
\label{eq:F-L}
\end{equation}
where $\vek(\omega)$ is
\begin{equation}
\vek(\omega)
\equiv 
1 +\infint dp\frac{\hat F(k)}{i( -\omega + kp)}\pdl{f_0}{p}(p).
\label{eq:vek}
\end{equation}

\subsection{Dispersion relation}
From the inverse-Laplace transform of  Eq.(\ref{eq:F-L}),
\begin{align}
  \delta f_k(t) &= \int_{-\infty-i\sigma}^{\infty-i\sigma}
e^{i\omega t}\tdk(\omega)\frac{d\omega}{2\pi} \\
&=
 \int_{-\infty-i\sigma}^{\infty-i\sigma}
e^{i\omega t}
\frac{1}{\vek(\omega)}\frac{-1}{{i( -\omega + kp)}}\widehat{\delta f}(k,p,0)
\frac{d\omega}{2\pi}.
\end{align}
Now we continuously move the integration contour to 
               upper half of complex $\omega$ plane while avoiding 
               the singular points.
Then the contributions from except of the pole can be neglected,
because of the factor $\exp(i\omega t)$ ($Im(\omega)>0$).

Then the growth rate of each mode of the fluctuation is obtained
by the solution of the equation 
\begin{equation}
   \label{eq:dsp-0}
\vek(\omega)=0.   
 \end{equation}
Eq.(\ref{eq:dsp-0}) is the dispersion relation.
If Eq.(\ref{eq:dsp-0}) has the solution where
the inequality $Im(\omega)<0$ is satisfied, the fluctuation is unstable.

\subsection{Dynamical stability of systems with power-law potentials}

Now we consider the case that the potential is power-law, the
pair-potential $U$ is
\begin{equation}
U(x) = A|x|^\alpha.
\end{equation}
Assuming the interaction is attractive, $A>0$.
$\alpha=1$  for ``sheet model''.
The Fourier-transformed potential is~\cite{Lighthill}
\begin{align}
\hat U(k) &\equiv \infint dx e^{ikx} A|x|^\alpha\\
&= 2A\left\{
-\left(\sin\frac{\alpha \pi}{2}\right) 
\frac{\Gamma(\alpha+1)}{|k|^{\alpha+1}}
\right\} \ \ (\alpha \ne 0,2,4,\cdots , -1,-3,\cdots).
\label{eq:disp-1}
\end{align}
For simplicity, we choose the stationary state as
\begin{equation}
\ds f_0 = n_0 \delta(p),
\label{eq:initial}
 \end{equation}
where $n_0$ is the number density of particles.

The dispersion relation is
\begin{equation}
\vek(\omega)=1 +  2n_0A \left\{
\left(\sin\frac{\alpha \pi}{2}\right) 
\frac{\Gamma(\alpha+1)}{|k|^{\alpha-1}}
\right\}\frac{1 }{\omega^2} =0,
 \end{equation}
and $\omega$ which satisfy the dispersion relation is
\begin{equation}
\omega ^2=-2n_0A \left\{
\left(\sin\frac{\alpha \pi}{2}\right) 
\frac{\Gamma(\alpha+1)}{|k|^{\alpha-1}}
\right\}.
\label{eq:dispersion-relation}
 \end{equation}
When $\omega ^2<0$, the system is unstable. 
Eq. (\ref{eq:dispersion-relation}) can be reduced to
\begin{equation}
  \omega = \pm i\sqrt{
2n_0A \left\{
\left(\sin\frac{\alpha \pi}{2}\right) \Gamma(\alpha+1)\right\}
}\,\cdot|k|^{(1-\alpha)/{2}}.
\label{eq:dispersion-relation2}
 \end{equation}
From Eq.(\ref{eq:dispersion-relation2})
we can classify the evolution of the perturbation into three types;
(i) when $0<\alpha<1$, the growing rate increase monotonously for
$|k|$.
(ii) when $\alpha=1$, the fluctuations for all scale grows at same
rate.
(iii) when $1<\alpha<2$, the growing rate decrease monotonously for
$|k|$.

The cold-random condition corresponds to 
the mixed states of fluctuation with all wavelength,
so called ``white noise''.
The larger the value of $\alpha$ is, the smaller
the growth rate of the fluctuation with large wavevector is.
This is consistent with the numerical results in Sec.
\ref{sec:simulation}
that a large whirlpool is
formed in $\mu$ space in the case for the large value of $\alpha$.
 
\section{Concluding remark}
\label{sec:summary}
In this paper we have studied structure formation of many-body systems
with power-law potentials.
Firstly we have investigated structure formation in this model
by numerical simulation. As results, we have found that 
behaviors of time evolution
are different depending on the power index of the potential.
When $\alpha$ is slightly above 1, fractal structure is formed similar to 
the sheet model \cite{hktk-1}. 
On the other hand, when $\alpha\gtrsim 1.5$, fractal structure is not formed.

In order to explain these numerical results, we have also analyzed linear
perturbation of collisionless Boltzmann equation
 to derive the dispersion relation which represent
the growth rate of each mode of the fluctuation.
As results, we have found that for large value of $\alpha$ the growth rate
of small scale is suppressed.
In addition we can explain the sheet model ($\alpha=1$) is marginal in the sense
all scale fluctuations grow at same rate.
There is a slight difference between 
the initial condition used in the numerical simulation and
the stationary state employed in the perturbation,  
that is, we set $N$ particles in finite region for simulation,
whereas the unperturbed state $f_0$ is  infinitely spread. 
Nontheless we think the linear perturbation grabs the core of the
instability, especially for short time scale.

In our simulation the fractal structure is formed through hierarchical
clustering. That is, clusters are created first in small spatial scale,
then grow to large scale \cite{hktk-2}.  
The spatial randomness in the initial condition and 
finiteness of the number of particles $N$ imply that
relative fluctuation in mass density is large for small spatial scale.
This enhancement of initial fluctuation in small spatial scale  is 
probably the ``seed'' of spatial structure formation.

Fractal properties of the structures can differ from place to place 
of the system. Most fractal structures are not exactly self-similar but
can contain various inner structures.
Using generalized dimension $D_q$ ~\cite{Halsey-mf-pra-1986} 
one can unveil details of fractal structures. 
Here we show the generalized dimension of the spatial structure formed 
in the power-law potential model
with $\alpha=1.125$ in Fig.\ref{fig:multi}.
We find that the generalized dimension is almost constant ($D_q\simeq
0.83$)
for some range of $q$ at $t=5.0$ (Fig.\ref{fig:multi}, top).
Therefore  we conclude that the fractal structure formed is 
{\it mono-fractal} at $t=5.0$ as far as we observed.

As discussed in our previous paper \cite{hktk-3}, 
however, the fractal structure 
is a transient state and relaxes finally.
We find that the 
multifractal nature emerges due to the relaxation at late times (the bottum of
Fig. \ref{fig:multi}).
The boxcounting dimension $D_0$ has relaxed to almost $1$, while the correlation
dimension $D_2$ has not yet (the bottom of Fig.\ref{fig:multi}).
In other words, the relaxation of the boxcounting dimension $D_0$ is much faster than the
correlation dimension $D_2$. The detailed analysis will be a future
work.

We noted that scale free property of the potential may be 
a key-stone to understand the dynamical mechanism of the 
emergence of fractal spatial structure.
In this paper we have clarified that scale free property
does not immediately imply fractal structure. 
Potentials with different power-index makes different temporal behaviors,
and the gravitational system ( the sheet model ) is 
a special case in all the 1-dimensional systems with
scale-free potential.

As relevant works, structure frormation in other one-dimensional
self-gravitating systems in an expanding universe has been
studied~\cite{RFN,Miller-2}. 
They claimed that their system is not a simple fractal or even a
regular multi-fractal, but bifractal~\cite{Miller-2}.
Indications of this behavior have also been found for the ``quintic''
model~\cite{AFMG}.
It is a future work to clarify the relevance between the models comprehensibly.
In addision, the dependence of the spatial dimension on the structure formation 
will be also interesting and important subject of future.

\acknowledgements
We would like to thank Angelo Vulpiani for useful comments.
H.K. is supported by JSPS Fellowship for Young Scientists.
T.K. is supported by Grant-in-Aid for Scientific Research
from the Ministry of Education,Culture,Sports,Scienceand Technology.

\appendix
\section{proof of Eq.(\ref{eq:F-L})}
In this appendix we lead Eq.(\ref{eq:F-L}) by the Fourier-Laplace
transform of Eq.(\ref{eq:cbe-l}).
In terms of the transform (\ref{eq:F-L}), Eq. (\ref{eq:cbe-l})
become following; The first term of the left hand side of
Eq. (\ref{eq:cbe-l}) is transformed to
\begin{align}
\int_0^\infty dt e^{-i\omega t}\infint dx e^{ikx}  \pdl{}{t}  \delta f(x,p,t)
&=
\left[ e^{-i\omega t} \widehat{\delta f}(k,p,t)\right]_{t=0}^{t=\infty}
-
(-i\omega)\int_0^\infty dt e^{-i\omega t} \widehat{\delta f}(k,p,t)
\\
&=
-\widehat{\delta f}(k,p,t=0)
+i\omega \tdf(k,p,\omega),
\end{align}
here $\omega$ is a complex number and $Im \omega<0$.
The second term of the left hand side of
Eq. (\ref{eq:cbe-l}) is transformed to
\begin{align}
\int_0^\infty dt e^{-i\omega t}\infint dx e^{ikx}  p\pdl{}{x}  \delta f(x,p,t)
&=
- ikp\int_0^\infty dt e^{-i\omega t}
\widehat{\delta f}(k,p,t)  
= -ikp\tdf(k,p,\omega).
\end{align}
Then the left hand side of Eq. (\ref{eq:cbe-l}) is transformed to
\begin{equation}
(lhs)= -i(-\omega +kp)\tdf(k,p,\omega) - \tdf(k,p,0).
\end{equation}
On the other hand,
the right hand  side of Eq. (\ref{eq:cbe-l}) is transformed to
\begin{align*}
(rhs)&= 
-\int_0^\infty dt e^{-i\omega t} \infint dx e^{ikx}
\left\{
\infint dx' F(x-x') \infint dp' \delta f(x',p',t) \pdl{}{p}f_0(p)
\right\}\\
&=
\infint dy e^{iky}F(y) \infint dp' \tdf(k,p',\omega) 
\pdl{f_0}{p}(p) \\
&= \hat F(k) \infint dp' \tdf(k,p',\omega) \pdl{f_0}{p}(p),
\end{align*}
where
\begin{equation}
 \tilde{\delta f}(x',p',\omega)
\equiv \int_0^\infty dt e^{-i\omega t} \delta f(x',p',t).
\end{equation}

Eq. (\ref{eq:cbe-l}) is transformed to
\begin{align}
\tdf(k,p,\omega)
&= \frac{-\hat F(k)}{i(-\omega + kp)}\pdl{f_0}{p}(p)
\cdot \left(\infint dp' \tdf(k,p',\omega)\right) +
 \frac{-1}{{i(-\omega + kp)}}\widehat{\delta f}(k,p,0),
\label{eq:cbe-l2}
\end{align}
where
\begin{align*}
  \tdk(\omega)\equiv \infint dp \tdf(k,p,\omega).
\end{align*}
Integrating both side of Eq.(\ref{eq:cbe-l2}) by $p$, we obtain
\begin{align}
\left(1 +\infint dp\frac{\hat F(k)}{i( -\omega + kp)}\pdl{f_0}{p}(p) \right)
\tdk(\omega)
&= \frac{-1}{{i( -\omega + kp)}}\widehat{\delta f}(k,p,0).
\label{eq:cbe-l3}
\end{align}
Using Eq.(\ref{eq:vek}), we obtain Eq. (\ref{eq:F-L}).

\begin{figure}[H]
 \caption{An example of fractal structure is formed of the model (\ref{eq:model-sheet}) from non-fractal structure.
The number of particles is $N=2^{15}$.
The left figure represents the $\mu$ space at $t=9.375$. The right figure
represents the box-counting dimension.}
    \label{fig:corr-and-xu-rnd-003}
  \end{figure}

\begin{figure}[H]
  \begin{center}
    \caption{The snap-shots of the $\mu$ space for $\alpha =1.125$.
$N=65536$.
      Time are 
$t=4.4$, $5.0$, $9.4$ from the top to the bottom.}
    \label{fig:11mu}
  \end{center}
\end{figure}

\begin{figure}[H]
  \begin{center}
    \caption{ Box counting dimension $D$ of the spatial distribution for $\alpha =1.125$
and $t=5.0$. The plus symbols represent our data, and solid and dashed
lines correspond
to $D=0.83$ and $D=1$, respectively.}
    \label{fig:11xi}
  \end{center}
\end{figure}

\begin{figure}[H]
  \begin{center}
    \caption{The snap-shots of the $\mu$ space for $\alpha =1.5$.
$N=65536$.
Time are $t=11.0$, $15.7$ and $23.6$ from the top to the bottom.
}
    \label{fig:15mu}
  \end{center}
\end{figure}

\begin{figure}[H]
  \begin{center}
\caption{
Box counting dimension $D$ of the spatial distribution for
   $\alpha =1.5$ and $t=23.6$.}
\label{fig:15xi}
  \end{center}
\end{figure}

\begin{figure}[H]
  \begin{center}
    \caption{The snap-shot of the $\mu$ space for $\alpha=1.125$ (left).
The generalized dimension $D_q$ (right).
Time are $t=5.0$, $9.4$ and $37.7$ from the top to the bottom.}
    \label{fig:multi}
  \end{center}
\end{figure}

\begin{thebibliography}{10}

\bibitem{Fersht-1998}
A. Fersht, {\em Structure and Mechanism in Protein Science},
W.H.Freeman, New York (1998).
%

\bibitem{Karplus-2000}
M. Karplus, {\em J. Phys. Chem.} {\bf B 104}, 11 (2000).


\bibitem{Peebles-book-1980}
P. J. E. Peebles, {\em The Large Scale Structure of the Universe},
Princeton University Press (1980).


\bibitem{dauxois-2000}
T.~Dauxois, P.~Holdsworth and S.~Ruffo,
\newblock {\em European Physical Journal} {\bf B 16}, 659 (2000).

\bibitem{barre-2001}
J. Barr\'{e}, T.~Dauxois and S.~Ruffo,
\newblock {\em Physica} {\bf A 295}, 254 (2001).

\bibitem{hktk-1}
H. Koyama and T. Konishi,
\newblock {\em Phys. Lett.} {\bf A 279}, 226 (2001).

\bibitem{hktk-2}
H. Koyama and T. Konishi,
\newblock {\em Euro. Phys. Lett.} {\bf 58}, 356 (2002).

\bibitem{hktk-3}
H. Koyama and T. Konishi,
\newblock {\em Phys. Lett.} {\bf A 295}, 109 (2002).




\bibitem{MPT}Lj. Milanovi\'{c}, H.A. Posch and W. Thirring, 
{\em Phys. Rev.} {\bf E 57}, 2763 (1998).







\bibitem{hohl-feix-1967}
F. Hohl and M.R. Feix,
\newblock {\em The Astrophysical Journal} {\bf 147}, 1164 (1967).


\bibitem{rybivki-1971-apss}
G.B. Rybicki,
\newblock {\em Astrophysics and Space Science} {\bf 14},
56 (1971).

\bibitem{severne-1984-apss}
M. Luwel, G. Severne, and P.J. Rousseeuw,
\newblock {\em Astrophysics and Space Sciences} {\bf 100}, 261 (1984).


\bibitem{Miller-2}
B.N. Miller and C.J. Reidl Jr.,
\newblock {\em The Astrophysical Journal} {\bf 348}, 203 (1990).


\bibitem{Galactic-Dynamics}
J. Binney and S. Tremaine,
\newblock {\em Galactic Dynamics}, Princeton Univ. Press (1987).

\bibitem{Inagaki-Konishi-1993} 
S. Inagaki and T. Konishi,
\newblock{\em Publ.  Astron. Soc. Jpn.} {\bf 45}, 733 (1993).

\bibitem{Lighthill} 
M.J. Lighthill,
\newblock{\em An Introduction to Fourier Analysis and 
Generalised Functions}, Campridge Univ. Press (1958).

\bibitem{Halsey-mf-pra-1986}
T. C. Halsey, M.H. Jensen, L.P. Kadanoff, I. Procaccia and
	B.I. Shraiman,
Phys. Rev. A {\bf 33} 1141--1151 (1986) 

\bibitem{RFN}J.-L. Rouet, M.R.Feix and M.Navet,
{\em Fractals in Astronomy}, A. Heck, Vistas in Astronomy,
357 (1990)
\bibitem{AFMG}E. Aurell, D Fanelli, and P. Muratore-Ginanneschi,
Physica D, {\bf 148} 272--288 (2001)


\end{thebibliography}
\end{document}